\documentclass[journal]{vgtc}                     % final (journal style)
\onlineid{1317}

\vgtccategory{Research}

\vgtcpapertype{system}

\title{Robust Dual-Modal Speech Keyword Spotting for XR Headsets}

\author{%
  \authororcid{Zhuojiang Cai}{0009-0005-3404-118X},
  \authororcid{Yuhan Ma}{0009-0008-4679-0454}, and 
  \authororcid{Feng Lu}{0000-0001-9064-7964}, Senior Member, IEEE
}

\authorfooter{
  \item
  	Zhuojiang Cai, Yuhan Ma, and Feng Lu are with State Key Laboratory of Virtual Reality Technology and Systems, School of Computer Science and Engineering, Beihang University, Beijing, China.
  	E-mail: \{caizhuojiang\,$|$\,raphael.mayuhan\,$|$\,lufeng\}@buaa.edu.cn\,.
  \item
        Zhuojiang Cai and Yuhan Ma contribute equally to this work. Feng Lu is the corresponding author.
}

\abstract{%
  While speech interaction finds widespread utility within the Extended Reality (XR) domain, conventional vocal speech keyword spotting systems continue to grapple with formidable challenges, including suboptimal performance in noisy environments, impracticality in situations requiring silence, and susceptibility to inadvertent activations when others speak nearby. These challenges, however, can potentially be surmounted through the cost-effective fusion of voice and lip movement information. Consequently, we propose a novel vocal-echoic dual-modal keyword spotting system designed for XR headsets. We devise two different modal fusion approches and conduct experiments to test the system's performance across diverse scenarios. The results show that our dual-modal system not only consistently outperforms its single-modal counterparts, demonstrating higher precision in both typical and noisy environments, but also excels in accurately identifying silent utterances. Furthermore, we have successfully applied the system in real-time demonstrations, achieving promising results. The code is available at \url{https://github.com/caizhuojiang/VE-KWS}.
}


\teaser{
  \centering
  \includegraphics[width=\linewidth]{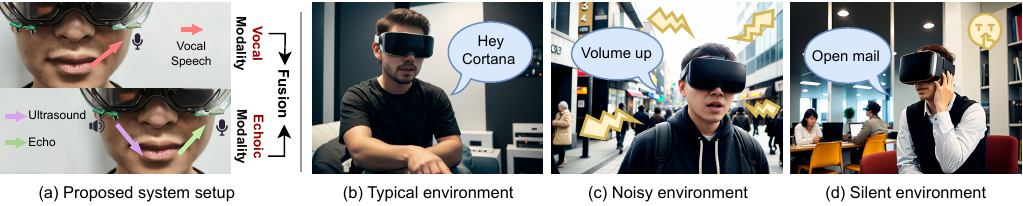}
  \caption{%
  	Our dual-modal system significantly extends the available scenarios for speech interaction with XR headsets. (a) We propose a vocal-echoic dual-modal keyword spotting system. (b) In typical environments, such as working or entertaining indoors, dual-modal system performs accurately and can block out the speech of speakers nearby. (c) In noisy environments, such as on the street or in the subway, our dual-modal method outperforms their single-modal counterparts. (d) When unable or unwilling to make a voice, for example, when others are working or sleeping nearby, our system continues to function effectively.%
  }
  \label{fig:teaser}
}

\graphicspath{{figs/}{figures/}{pictures/}{images/}{./}} % where to search for the images

\usepackage{tabu}                      % only used for the table example
\usepackage{booktabs}                  % only used for the table example
\usepackage{lipsum}                    % used to generate placeholder text
\usepackage{mwe}                       % used to generate placeholder figures

\usepackage{mathptmx}                  % use matching math font

\usepackage{amsmath}

\begin{document}

%%%%%%%%%%%%%%%%%%%%%%%%%%%%%%%%%%%%%%%%%%%%%%%%%%%%%%%%%%%%%%%%
%%%%%%%%%%%%%%%%%%%%%% START OF THE PAPER %%%%%%%%%%%%%%%%%%%%%%
%%%%%%%%%%%%%%%%%%%%%%%%%%%%%%%%%%%%%%%%%%%%%%%%%%%%%%%%%%%%%%%%

\firstsection{Introduction}

\maketitle

%% \section{Introduction} %for journal use above \firstsection{..} instead
In recent years, Extended Reality (XR) headsets, including Virtual Reality (VR), Augmented Reality (AR), and Mixed Reality (MR) headsets, have gained widespread attention. These devices, serving as bridges between the virtual and real worlds, have the potential to profoundly reshape how people live, work, and entertain themselves. With the increasing popularity of MR and AR headsets like the Apple Vision Pro and Microsoft HoloLens, XR headsets are finding utility in an ever-expanding array of scenarios.

Efforts to develop natural and effective user interaction methods with XR headsets have been a critical research focus. A growing number of headsets, including the Apple Vision Pro, have shown a preference for intuitive interaction methods such as speech, gestures, and gaze. These methods, in comparison to tools like controllers, are more readily accepted and user-friendly. Among them, speech interaction can be used for menu selection, locomotion control, and combined with gaze for object movement, etc. However, the extensive use of speech interaction has been hindered by limitations in speech recognition methods, including Automatic Speech Recognition (ASR) and Speech Keyword Spotting (KWS), which lack robustness across various real-world environments.

Conventional vocal keyword spotting methods encounter various challenging usage scenarios. Specifically, we have identified three common and demanding situations: 1) In noisy environments such as bustling streets, crowded malls, or noisy public transport, keyword spotting accuracy significantly degrades. 2) Users often cannot vocalize when others are working or resting nearby, or due to privacy concerns and social awkwardness. In such cases, vocal keyword spotting fails entirely. 3) Vocal speech keyword spotting systems are susceptible to interference and even false triggering when others are speaking nearby. 

These challenges highlight the limitations of traditional vocal keyword spotting approaches and emphasize the need for more robust and versatile solutions. Addressing the second challenge, significant progress has been made in the field of silent speech interfaces. Various types of data, including facial imagery\cite{fung_end--end_2018, xu_lcanet_2018}, ultrasonic imaging\cite{kimura_sottovoce_2019}, EMG\cite{lee_emg-based_2008, kapur_alterego_2018, wang_silent_2020}, motion sensing\cite{rekimoto_derma_2021}, strain sensing\cite{kunimi_e-mask_2022}, and other sensing forms have been investigated for recognizing silent speech. Furthermore, some research has leveraged the transmission and reception of ultrasonic waves to detect mouth movements\cite{gao_echowhisper_2020, zhang_celip_2021, zhang_echospeech_2023}, offering contactless and cost-effective solutions. EchoSpeech\cite{zhang_echospeech_2023}, for instance, employed Frequency-Modulated Continuous Wave (FMCW) and demonstrated impressive silent speech spotting capabilities using off-the-shelf speakers and microphones mounted on glasses frames. However, this work did not explore the use of low-frequency vocal speech information from microphone audio, which could extend its application to a broader range of vocal speech recognition scenarios, thereby addressing the remaining two challenges.

Recognizing that vocal speech and lip movement can contribute from different angles to the comprehension of a speaker's speech, and drawing inspiration from related research in audio-visual speech recognition\cite{pomianos_recent_2003, ding_audio-visual_2018, zhou_modality_2019, shi_robust_2022} that supports this idea, we posit that vocal speech information and mouth movement information obtained through ultrasonic echo could similarly offer diverse perspectives on a speaker's speech. The integration of these two modalities may provide a wealth of prior knowledge for speech recognition, potentially leading to improved performance.

Therefore, this paper introduces a novel vocal-echoic dual-modal keyword spotting method for XR headsets. By combining vocal speech and ultrasonic echo features, it achieves robust keyword spotting in various scenarios. We designed and implemented this dual-modal keyword spotting system on Microsoft HoloLens 2 with custom hardware. We devised two different modal fusion approaches and verified their effectiveness through experiments. Furthermore, we compared the dual-modal system with single-modal systems in low-noise environments, diverse noisy environments, and scenarios with interference from other speakers. The results demonstrate that our proposed dual-modal system consistently outperforms its single-modal counterparts in the majority of scenarios, without compromising silent speech spotting performance. Finally, we applied the system in practical examples, achieving promising results. Our work significantly broadens the scope of speech keyword spotting applications.

In summary, our contributions can be outlined as follows:
\begin{itemize}
    \item We propose a vocal-echoic dual-modal speech keyword spotting (KWS) system, enabling robust speech recognition in a broader range of scenarios.
    \item We conducted an ablation study to design a lighter CNN model for the echoic modal KWS, reducing its demand for computing resources in the headset.
    \item We conducted experiments to assess the system's performance in various noisy environments, situations with nearby speakers, and silent scenarios, demonstrating that our dual-modal system outperforms traditional vocal systems in all of these scenarios.
\end{itemize}

\section{Related Work}

\subsection{Vocal Speech Keyword Spotting}

Keyword Spotting (KWS) is the task of detecting a predefined set of keywords from an audio stream. Compared to Automatic Speech Recognition (ASR), KWS can be deployed on edge devices with low computational requirements and without the need for cloud connectivity, eliminating privacy concerns. Therefore, it finds extensive applications in various domains such as user interface interactions, smart home control, and command triggers.

Typically, KWS refers to the detection of vocal speech. However, in this paper, we also employ a silent speech keyword spotting using echoic modality. To differentiate, we refer to the conventional KWS as vocal speech keyword spotting in this context.

\textbf{ASR-based KWS.} Some studies utilize ASR systems to convert speech signals into text and then identify keywords through text matching techniques\cite{karpov_low-resource_2017, michaely_keyword_2017, rosenberg_end--end_2017}. While this approach eliminates the need for specialized training of predefined keywords, offering greater flexibility, it inherits the drawbacks of ASR, including computational demands and privacy concerns.

\textbf{HMM-based KWS.} The KWS system based on Hidden Markov Models (HMM) was proposed three decades ago\cite{rose_hidden_1990, wilpon_improvements_1991, rohlicek_continuous_1989}. In this approach, speech samples of each keyword are used to train the corresponding HMM for that keyword, and non-keyword (filler) speech segments are used to train a filler HMM. At runtime, the input audio stream is matched against these HMM models. The Viterbi algorithm is commonly employed to find the most likely state sequence. If the matching probability exceeds a predefined threshold, the system identifies the input as containing the keyword. Although these systems perform well, executing multiple HMM model matches with the Viterbi algorithm still requires significant computational power\cite{chen_small-footprint_2014, lopez-espejo_deep_2022}.

\textbf{Deep KWS.} In the past decade, the rapid advancement of deep learning has led to extensive research and application of deep KWS approches\cite{lopez-espejo_deep_2022}. These advancements have resulted in reduced computational complexity and improved performance of KWS systems. Common architectures, such as fully-connected networks\cite{chen_small-footprint_2014} and Recurrent Neural Networks (RNNs)\cite{zhuang_unrestricted_2016, sun_max-pooling_2016, kumar_convolutional_2018}, have been studied and proven effective in KWS systems. Inspired by computer vision research, KWS based on Convolutional Neural Networks (CNNs)\cite{sainath_convolutional_2015, tang_deep_2018, choi_temporal_2019, xu_depthwise_2020, kim_broadcasted_2023} has garnered significant attention due to its straightforward architecture and ease of tuning\cite{tang_deep_2018}. Notably, Tang and Lin\cite{tang_deep_2018} applied the concept of residual learning to KWS, creating high-performance models with a small footprint. Subsequently, TC-ResNet\cite{choi_temporal_2019} replaced the original convolutional modules in residual blocks with temporal convolutions, and DC-ResNet\cite{xu_depthwise_2020} introduced depthwise separable convolutions. These innovations reduced the parameter count and computational complexity. Kim et al.\cite{kim_broadcasted_2023} introduced a novel network architecture called Broadcast Residual Learning, which consistently outperforms previous models with an equivalent parameter count. However, despite demonstrating excellent performance on public datasets, these methods experience significant performance degradation in noisy environments\cite{lopez-espejo_novel_2021, prabhavalkar_automatic_2015, cioflan_towards_2022}. This issue will be thoroughly investigated in this paper.

\subsection{Silent Speech Interface on Wearables}

Silent Speech Interface (SSI) can be viewed as a generalized form of speech recognition. It involves capturing non-vocal information using various sensors to recognize speech utterances from users.

\textbf{Contacting SSI.} In wearable devices, SSIs implemented through diverse sensor technologies have been widely researched. Some approaches place magnetometers\cite{cheah_wearable_2018, hofe_small-vocabulary_2013, bedri_toward_2015} or capacitive sensors\cite{kimura_mobile_2021, li_tongueboard_2019} inside the mouth to capture movements of the mouth and tongue, reconstructing speech utterances. Others employ sensors attached closely to the skin, recognizing speech through forms of information like ultrasound imaging\cite{kimura_sottovoce_2019}, Electromyography (EMG)\cite{lee_emg-based_2008, kapur_alterego_2018, wang_silent_2020}, motion sensors\cite{rekimoto_derma_2021}, or stress sensors\cite{kunimi_e-mask_2022}. However, these methods require skin contact or even intrusion into the mouth, which may be uncomfortable for users\cite{zhang_echospeech_2023}.

\textbf{Contact-free SSI.} Contact-free SSIs offer a more comfortable user experience, leading to increased research interest. Some efforts attempted to install cameras on headphones\cite{chen_c-face_2020} or neck-mounted devices\cite{kimura_tielent_2020, zhang_speechin_2021}; however, these methods faced challenges related to high power consumption and privacy risks. Additionally, some works utilize active acoustic methods, employing microphones and speakers on smartphones\cite{gao_echowhisper_2020}, VR headsets\cite{zhang_celip_2021}, and glass-frames\cite{zhang_echospeech_2023} to achieve low-power, high-performance silent speech keyword spotting. Among them, CELIP\cite{zhang_celip_2021} requires a setup directly facing the user's mouth, including a pair of relatively large microphones and speakers, whereas EchoSpeech\cite{zhang_echospeech_2023} integrates compact components under the lower frame of glasses, ensuring a less obtrusive design. 

Although these contact-free methods achieve silent speech recognition with low power consumption and high performance, their potential to integrate with vocal keyword spotting systems for enhanced performance and broader application scenarios remains unexplored.

\begin{figure*}[]
 \centering % avoid the use of \begin{center}...\end{center} and use \centering instead (more compact)
 \includegraphics[width=\linewidth]{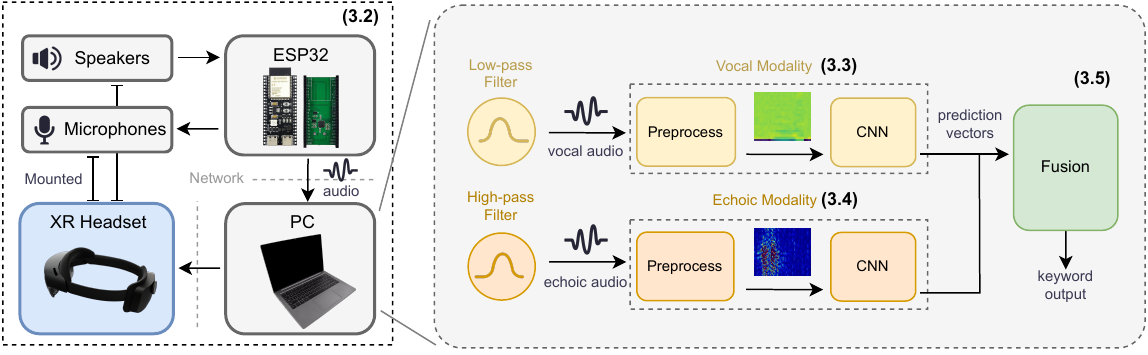}
 \caption{Overview of vocal-echoic dual-modal KWS system for XR headset. (left) Hardware diagram of experimental equipment. The speakers and microphones are mounted on the XR headset, connected to an ESP32, which sends the audio to a PC over the network. The PC is used for algorithm implementation and experiments, and it sends the detected keywords back to the application on the headset. (right) Algorithm flowchart. The audio is separately filtered and input into the vocal and echoic modal KWS pipelines. The predicted vectors obtained from these pipelines are then fed into the fusion module to generate the keyword output.}
 \label{fig:system-design}
\end{figure*}

\subsection{Speech Interaction in XR}

Speech interaction allows users to control devices in XR through verbal commands, either independently or in collaboration with other interaction methods like controller, gesture, and gaze. This approach provides a more flexible and user-friendly way to operate devices in XR.

\textbf{Hand-free Interaction.} The use of speech commands for hands-free interaction is a common practice in XR\cite{hombeck_tell_2023, rantamaa_evaluation_2022, grinshpoon_hands-free_2018}. For instance, voice commands are utilized to control locomotion in VR without the need for hand gestures\cite{hombeck_tell_2023}. In dental implant surgeries where hands might be occupied, speech commands can replace menu clicks in AR\cite{rantamaa_evaluation_2022}. Studies suggest that voice-based interactions can be as efficient as gesture-based interactions\cite{lee_usability_2013} and free up hands for other tasks\cite{monteiro_hands-free_2021}.

\textbf{Multimodal Interaction.} Other studies have explored interactive methods combining speech commands with gesture and gaze. For example, using gestures to select a target and then executing operations through speech commands\cite{piumsomboon_grasp-shell_2014} is considered more efficient and accurate than relying solely on gestures\cite{lee_usability_2013}. Another approach involves using gaze and speech to control the movement of objects\cite{elepfandt_move_2012, kaur_where_nodate}. Wang et al.\cite{wang_interaction_2021} investigated a multimodal interaction method in AR that combines speech, gaze and gesture and demonstrated its superior efficiency compared to single-modal and dual-modal methods.

\section{Dual-Modal Keyword Spotting System}

\subsection{System Overview}

Our dual-modal keyword spotting (KWS) system represents an elegant enhancement of its commonly used single-modal vocal counterpart. It maintains audio streams as the sole input and predicted keywords as the output. 

Unlike traditional systems that directly use input audio for vocal KWS, our system separates audio into vocal and echoic modalities using bandpass filters. The echoic audio originates from ultrasonic waves emitted by speakers and reflected off the user's skin near the mouth. These two segments of audio are then separately processed by the vocal and echoic modal KWS pipelines before their predictions are fused into a single output.

The difference in frequency ranges between vocal and ultrasonic echoic audio makes this system feasible. Research has shown that the fundamental frequency (F0) of vocal speech typically falls within the range of approximately 100-240Hz. Even considering the harmonic information that may be present in vocal speech, the upper limit of the frequency bands used in conventional speech recognition Mel-filter banks is usually around 5000Hz. In contrast, the ultrasonic waves used in our system operate at frequencies above 17kHz. Hence, both modalities of audio can be concurrently captured using the same microphone, providing representations of two modalities for the same speech.

Both modal KWS pipelines in the system employ lightweight deep learning approaches. For vocal KWS, many methods have already achieved excellent results in typical low-noise scenarios. In this paper, we directly utilize one of the state-of-the-art convolutional neural networks (CNNs) for vocal KWS, which will be discussed in \cref{sec:vocal-modality}. For echoic KWS, EchoSpeech\cite{zhang_echospeech_2023} uses a ResNet18 as the backbone network. We significantly reduced the network's parameters with negligible performance degradation through an ablation study, enabling it to run with lower power consumption and latency. This will be discussed in \cref{sec:echoic-modality}.

The predictions from the two KWS pipelines are fused in the final stage of the system. We propose two fusion methods: reliability-based fusion and MLP-based fusion. Both have been experimentally shown to achieve higher accuracy than single-modal results. This demonstrates that our fusion methods can comprehensively consider different dimensions of information from the two modalities for the same speech, resulting in more accurate keyword spotting. These two fusion methods will be introduced in \cref{sec:fusion}.

\begin{figure}[tb]
 \centering
 \includegraphics[width=\columnwidth]{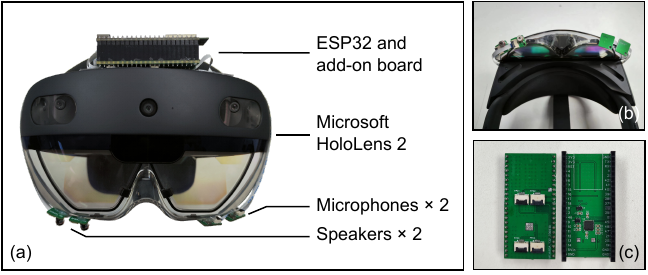}
 \caption{Hardware Setup. (a-b) Front view and bottom view of our implementation with HoloLens headset. (c) ESP32 Add-on board.}
 \label{fig:hardware}
\end{figure}

\subsection{Hardware Implementation}
\label{sec:harware}

Considering that the performance of the echoic modality is highly dependent on the positions of the speakers and microphones, it is worth noting that the positions of the components integrated into commercial XR headsets do not align optimally with the requirements of our system. 

Therefore, we employed a microcontroller board and a custom-designed add-on board to connect two pairs of small off-the-shelf speakers and microphones, which were mounted on Microsoft HoloLens 2 to implement the hardware for this system, as shown in \cref{fig:hardware} (a-b). This positional configuration has demonstrated strong performance in glasses-frames\cite{zhang_echospeech_2023} for echoic modal KWS. 

The microphones and speakers are both mounted on the lower edge of the lenses of HoloLens 2. Two speakers are along the lower edge of the right lens, while two microphones are along that of the left lens. The direction of the speakers and microphone holes is oriented downward. The distance between the speaker and the microphone closer to the nose is 7.2 cm, while the distance between the two speakers and between the two microphones is 1.8 cm. Additionally, The average vertical distance from the microphone plane to the horizontal midline of the mouth is 3.9 cm, measured across 15 participants wearing the device.

During the keyword spotting process, user utterances are captured by the microphones. Simultaneously, the speakers emit continuous ultrasonic waves, with a portion being reflected off the user's face and mouth, subsequently captured by the microphones as echoes. Hence, the signals received by the microphones can be separated into two components: vocal modality and echoic modality, enabling dual-modal keyword spotting and fusion.

In the experimental setup, an ESP32 development board is employed to control speaker playback, receive microphone signals, and transmit them to a computer or HoloLens for subsequent detection and fusion. The development board, speakers, and microphones are all common products purchased online. The development board used is the ESP32-S3-DevKitC-1-N8R8, the speakers are OWR-05049T-38D, and the microphones are ICS-43434. Furthermore, an add-on board has been designed to interface the modules with the ESP32 and decode audio, as shown in \cref{fig:hardware} (c).

\subsection{Vocal Modal KWS}
\label{sec:vocal-modality}

Keyword Spotting (KWS) in the vocal modality is a relatively mature technology, with a substantial amount of state-of-the-art work currently utilizing deep learning approaches\cite{tang_deep_2018, choi_temporal_2019, berg_keyword_2021, karpov_learning_2021, kim_broadcasted_2023}. These methods involve converting audio into Mel-frequency cepstral coefficients (MFCCs) and feeding the MFCCs into convolutional neural networks (CNNs) or other neural networks to produce keyword spotting results, which is the methodology our system adopts.

In our dual-modal KWS system, audio is first processed with a low-pass filter set to a 10 kHz cutoff frequency to isolate vocal audio. This audio is then subjected to a series of transformations, including pre-emphasis, framing, windowing, fast Fourier transform (FFT), Mel-frequency warping, logarithmic scaling, and discrete cosine transform, ultimately resulting in MFCCs features\cite{davis_comparison_1980, lopez-espejo_deep_2022}. These features are fed into a broadcast residual based CNN architecture\cite{kim_broadcasted_2023} to generate prediction vectors.

Differing from single-modal vocal KWS systems, where the \textit{argmax} in the prediction vector typically serves as the system output, we combine this vector with the one from the echoic modality KWS pipeline and input both vectors into a fusion module to obtain a unique prediction result.

\subsection{Echoic Modal KWS}
\label{sec:echoic-modality}

\begin{figure}[tb]
 \centering
 \includegraphics[width=\columnwidth]{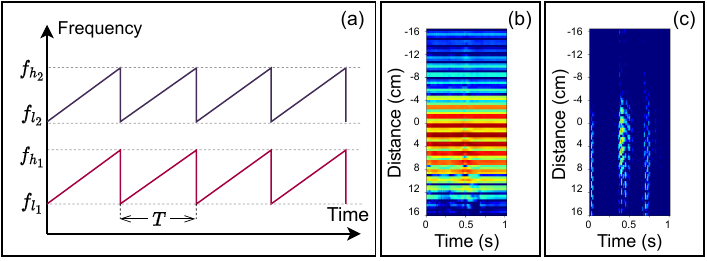}
 \caption{(a) The frequency-time diagram of the FMCW signals in the two frequency bands. (b-c) Original and differential Echo Profile.}
 \label{fig:echoic}
\end{figure}

The echoic modal KWS pipeline utilizes an active acoustic approach. Speakers installed at the bottom of the XR headset emit Frequency Modulated Continuous Waves (FMCW) in two frequency bands. Simultaneously, microphones, also positioned at the bottom, receive audio signals reflected off the skin near the mouth (echoes). By computing the cross-correlation between the echoes and the transmitted signals in each FMCW frame, features about mouth movements can be obtained, which can be learned by a Convolutional Neural Network (CNN) to achieve keyword spotting. 

\textbf{Why FMCW?} Different active acoustic methods, including Doppler effect, CIR, and FMCW, have been employed in prior works for silent speech interfaces. All of these methods have the potential to replace our echoic modality pipeline, as our fusion approach solely takes the keyword classification probability vectors as inputs. Among these methods, a previous study demonstrated the advantages of FMCW in capturing facial movements\cite{li_eario_2022}. Therefore, we have chosen this method to ensure optimal results after fusion.

\textbf{Transmitted signal.} Our transmitted signals are chirps in FMCW, where the frequency \(f\) linearly increases over time \(t\) within each chirp. This can be represented as \(f(t) = f_l + (f_h - f_l) \times t / T\), where \(f_l\) and \(f_h\) represent the lower and upper bounds of the frequency, and \(T\) represents the chirp period, as illustrated in \cref{fig:echoic} (a). In our system, two microphones simultaneously emit FMCW signals in two frequency bands: 17-20 kHz and 20.5-23.5 kHz, with a period of 12 ms. This frequency range is considered ultrasonic and is compatible with the sampling rates of commercial microphones and speakers, which operate at 48 kHz.

\textbf{Cross-correlation based FMCW.} The cross-correlation based FMCW method\cite{wang_c-fmcw_2018} is employed to process the echoes received by the microphone. By computing the cross-correlation between the echoes and the transmitted signals, the correlation at different sampling offsets within one chirp can be obtained. The sample shift is proportional to the distance from the microphone and speaker to the reflecting medium. Therefore, the cross-correlation reflects the magnitude of reflection at different distances, allowing for the resolution of 0.357 cm when the sampling rate is 48 kHz. This level of resolution enables the detection of positional changes of the skin near the user's mouth relative to the XR headset during speech.

\textbf{Echo Profile.} Calculating the cross-correlation for each consecutive chirp of the received signal produces a correlation graph with time on the horizontal axis and sample shift on the vertical axis, refer to as the Echo Profile\cite{li_eario_2022}. The differential Echo Profile, obtained by differencing the Echo Profile along the time axis, reveals the temporal movement characteristics of the mouth (see \cref{fig:echoic} (b-c)).

\textbf{CNN Model.} In echoic modal KWS, the differential Echo Profile serves as the input to the CNN network to predict keywords in the audio. Previous study\cite{zhang_echospeech_2023} utilized ResNet-18 as the backbone network, achieving good performance. However, ResNet-18 has significantly more parameters compared to the network used in vocal modality KWS, leading to additional computational demands. Therefore, we used ResNet-18 as the baseline and improved the architecture of the echoic modality network through ablation study. We achieved this by reducing the network width and replacing regular convolutional modules with depthwise separable convolutions, resulting in multiple models with different parameter counts. The ablation study is detailed in \cref{sec:exp-echoic-model}. We selected ResNet-18-1/4-DS as the network for our echoic KWS pipeline because it strikes a balance between performance and parameter count.

\subsection{Fusion Strategies}
\label{sec:fusion}

The fusion strategy comprises two objectives: on one hand, fully leveraging the distinct representations of user speech provided by both modalities to enhance prediction accuracy; on the other hand, remaining unaffected by unreliable information from one modality and utilizing trustworthy information from the other modality.

To achieve these goals, we explore two fusion methods: reliability-based fusion and MLP-based fusion. The former employs manually crafted features, calculating reliability indicators of prediction vectors from both modalities' pipeline outputs. These indicators are then adaptively used to determine fusion strategies and perform fusion operations. The latter employs a neural network approach, utilizing a Multi-Layer Perceptron model to learn the relationship between prediction vectors from both modalities and the fusion results.

\subsubsection{Reliability-based Fusion}

From the two objectives, as we aim to prevent the influence of unreliable information from individual modalities (for instance, the impact of vocal modality predicting "silence" during silent speech recognition), a natural approach is to assess the reliability of predictions from each individual modality. Therefore, we devised a reliability-based fusion strategy, adaptively fusing modalities based on reliability indicators.

\textbf{Reliability indicator.} We utilized a reliability index based on N-best log-likelihood difference and N-best log-likelihood dispersion. This fusion method, as demonstrated by Potamianos et al. \cite{pomianos_recent_2003}, has shown effectiveness in the realm of audio-video integration.

The first indicator is used to measure the class discrimination ability of each modality, while the second indicator supplements the additional N-best class likelihood ratios missing in the first indicator. The amalgamation of these two indicators allows for a more rational evaluation of the modality’s reliability. The definitions of the two indicators are as follows:

\textit{N-best Log-Likelihood Difference:}
\begin{equation}
     {L}_{m,t}=\frac{1}{N-1}\sum_{n=2}^{N}\log\frac{P\left(\mathbf{o}_{m,t}|c_{m,t,1}\right)}{P\left(\mathbf{o}_{m,t}|c_{m,t,n}\right)}
\end{equation}

\textit{N-best Log-Likelihood Dispersion:}
\begin{equation}
     {D}_{m,t}=\frac{2}{N\left(N-1\right)}\sum_{n=1}^{N}\sum_{n^{\prime}=n+1}^{N}\log\frac{P\left(\mathbf{o}_{m,t}|c_{m,t,n}\right)}{P\left(\mathbf{o}_{m,t}|c_{m,t,n^{\prime}}\right)}.
\end{equation}

Here, $P(o_{m,t}|c_{m,t,n})$ denotes the likelihood of observing the result $o_{m,t}$ given the class $c_{m,t,n}$. In these equations, $m$ and $t$ represent the modality and time respectively, and $n$ indicates the class ranking.

\textbf{Fusion process.} During the fusion process, reliability indicators are first employed to separately determine whether the prediction vectors from the two modalities are utilized. When both modalities are utilized, the reliability indicators are then used to calculate the fusion exponent. 

To determine the utilization of a modality, a threshold mechanism [$t_{l,m,t}$, $t_{d,m,t}$] has been implemented. A modality is considered reliable only if its reliability indicators meet specific threshold criteria. If one modality is deemed unreliable, we rely solely on the modality considered reliable to ensure the accuracy of the fusion results. This approach prevents interference from the unreliable modality. The boolean variable $R_{m,t}$ indicates whether the modalities exceed their respective thresholds, determining their reliability and whether their data will be included in the subsequent fusion process. The calculation of $R_{m,t}$ is defined as:

\begin{equation}
     {R}_{m,t}= ( {L}_{m,t}>{t}_{l,m,t})\land( {D}_{m,t}>{t}_{d,m,t}).
\end{equation}

In situations where both modalities are considered reliable, the information from both modalities should be fully utilized to obtain a more accurate result. In this case, we will reapply the reliability indicators to determine the fusion exponent $\lambda_{v,t}$: 

\begin{equation}
    \lambda_{v,t}=\frac{1}{1+\exp\left(-\sum_{i=1}^4\mathrm{w}_i\mathrm{a}_{i,t}d_{i,t}\right)}.
\end{equation}

Here, $d_t = [{L}_{v,t},  {D}_{v,t}, {L}_{e,t}, {D}_{e,t}]$ corresponds to the four reliability indicators of two modalities. ${w}_i$ assigns weights to the reliability indicators to ensure that the distinct reliability indicators of different modalities are appropriately aligned during the fusion process. It is important to emphasize that these weights are solely related to the vocal and echoic KWS models involved in the fusion and remains unaffected by changes in the data.

Furthermore, $\mathrm{a}_{i,t}$ signifies the reliability adjustment for two classes: ``silence'' and ``unknown''. Considering their auxiliary roles in the classification, when one modality provides ``silence'' or ``unknown'', it is imperative to take into full consideration the information from the other modality, thus necessitating a reduction in their reliability within a finite range. Differently, ``silence'' provided by the vocal modality should be accorded special attention. Typically, there is virtually no possibility of vocalization without any facial muscle movement. The value of $\mathrm{a}_{i,t}$ will only change to a specific value when there is at least one modality providing ``silence'' or ``unknown''; in all other cases, it remains a four-dimensional vector consisting of ones.

Subsequently, we weight the decisions of the two single-modal classifiers and utilize their log-likelihoods for linear combination, thereby obtaining the fusion results. Overall, our fusion process can be described by:
\begin{equation}
    \label{eq-fusion-cases}
    \begin{cases}
        P(\mathbf{o}_{f,t}|c) = P(\mathbf{o}_{v,t}|c)^{\lambda_{v,t}}P(\mathbf{o}_{e,t}|c)^{(1-\lambda_{v,t})}, & \mspace{10mu}  {R}_{v,t}\land \mspace{10mu}  {R}_{e,t} \\
        P(\mathbf{o}_{f,t}|c) = P(\mathbf{o}_{v,t}|c), & \mspace{10mu}  {R}_{v,t}\land\lnot {R}_{e,t} \\
        P(\mathbf{o}_{f,t}|c) = P(\mathbf{o}_{e,t}|c), & \lnot {R}_{v,t}\land\mspace{10mu} {R}_{e,t} \\
        \text{\emph{No credible result}}, & \lnot {R}_{v,t}\land\lnot {R}_{e,t}.
    \end{cases}
\end{equation}

\textbf{Determination of Parameters.} This fusion strategy initially requires the determination of eight parameters, four for setting thresholds and the other four for calculating exponents. This task employs a genetic algorithm to maximize the objective function that reflects the accuracy of the fusion results under different parameters. The remarkable global search capability of the genetic algorithm enables it to achieve satisfactory optimization effects under complex objective functions. 

In order to obtain a parameter vector of as high-quality as possible, we used the Latin Hypercube Sampling (LHS) initialization method to maximize coverage of the available parameter space for better search capability. The initial solution obtained with fewer iterations on a small dataset is used to replace the optimal solution in the initial population to cope with the high time cost of the genetic algorithm and its dependence on the initial population. 

In addition to determining the aforementioned eight parameters, it is imperative to ascertain the target values for parameter $\mathrm{a}_{i,t}$ when both modalities yield ``silence'' and ``unknown'' results separately. We need to determine four parameters, denoted as $\mathrm{a}_{v,s}$, $\mathrm{a}_{v,u}$, $\mathrm{a}_{e,s}$, and $\mathrm{a}_{e,u}$, corresponding to the modifications when each of the two modalities provides ``silence'' and ``unknown'' outcomes.Once the above eight parameters are established, we can employ a grid search to swiftly determine the values of these four parameters.

\subsubsection{MLP-based Fusion}

The MLP-based fusion method abandons manually crafted features and employs a straightforward Multi-Layer Perceptron (MLP) model to generate fusion results. It consists of an input layer, a hidden layer, and an output layer, with each layer fully connected to the next. The input vector is formed by concatenating the output vectors obtained from the vocal and echoic modal pipelines. This approch can be represented as follows:

\begin{equation}
    P(\mathbf{o}_{f,t}|c) = MLP([P(\mathbf{o}_{v,t}|c),P(\mathbf{o}_{e,t}|c)]).
\end{equation}

Here, \(P(\mathbf{o}_{f,t}|c)\) represents the fusion result at time \(t\) given class \(c\), \(P(\mathbf{o}_{v,t}|c)\) and \(P(\mathbf{o}_{e,t}|c)\) represent the prediction vectors for vocal and echoic modalities, respectively. The square brackets indicate the concatenation of the two output vectors.

We utilized the self-recorded dataset introduced in \cref{sec:data} to train the model. During training, this dataset was rigorously partitioned, with a strict demarcation between the data used for training and the data employed in subsequent experiments. Multiple data augmentation techniques were employed during training. For each data instance, four possible processing methods were applied: preserving clarity (no processing), introducing environmental noise interference, introducing vocal noise interference, and discarding a portion of vocal data. The environmental noise used in processing was extracted from the DEMAND\cite{thiemann_diverse_2013} noise dataset, as detailed in \cref{sec:experiment 2}. This noise data was then scaled to random signal-to-noise ratios before being combined with the training data. Additionally, vocal noise data was sourced from the Google Speech Commands\cite{warden_speech_2018} dataset, as explained in \cref{sec:experiment 4}. This vocal noise data was added to the training data after being multiplied by a fixed coefficient. Among these four processing methods, only one was randomly selected. Subsequently, the vector was multiplied by a random factor ranging between 0.95 and 1.05 to simulate random noise interference. Furthermore, the cross-entropy loss function and the Adam optimizer were employed during the training process.

In application, the system concatenates the prediction vectors from both the vocal and echoic modalities, and feed the resulting concatenated vector into the trained multi-layer perceptron. Subsequently, we select the class with the highest probability from the resulting prediction vector as the system's output.

\section{Experiments}

\subsection{Data}
\label{sec:data}

We recruited 15 participants (2 female and 13 males, from 20 to 27 years old) wearing our equipment and reading specific keywords to record audio data. Each participant read 30 repetitions of 10 comand words, and 5 repetitions of another 25 auxiliary words labeled as ``unknown'' to help distinguish unrecognized words. Additionally, each participant also wore the device to record data in which they did not speak and kept their mouth still, labeled as ``silence''. The data has a sampling rate of 48 kHz, allowing for the preservation of audio information with a maximum frequency of 24 kHz. As a result, the recorded data contains information from both vocal and echoic modalities.

The vocabulary used in the experiments is derived from the Google Speech Commands\cite{warden_speech_2018} dataset, which is the most commonly used open-source dataset for vocal KWS task. Therefore, it was natural for us to migrate the vocabulary from this dataset to our dual-modal KWS experiments. This allows our vocal modality model to be trained on a combination of this dataset, which has a large volume of data, and our proprietary dataset, maximizing its performance and robustness. This approach ensures that the comparative experimental results between single-modal and dual-modal performance are reliable.

\subsection{Evaluation Metric}

\textbf{Word Error Rate (WER).} The Word Error Rate (WER) is a metric used to evaluate the performance of speech recognition systems or natural language processing systems. It measures the difference between two texts, i.e., how many words in the predicted text are incorrect, missing, or redundant. The WER can be calculated as:

\begin{equation}
WER = \frac{S + D + I}{N} = \frac{S + D + I}{S + D + C}
\end{equation}

where \( S \) is the number of substitutions, \( D \) is the number of deletions, \( I \) is the number of insertions, \( C \) is the number of correct words, and \( N \) is the total word count in the text.

\subsection{Experiment 1: Echoic Model}
\label{sec:exp-echoic-model}

In this experiment, we conducted an ablation study on the echoic modality deep learning model using our proprietary dataset. The primary objective was to investigate the trade-off between model complexity (parameter count) and accuracy, with the ultimate goal of achieving a more streamlined and lightweight model.

\subsubsection{Experiment Setup}

We commenced our experiments with ResNet-18, a model that EchoSpeech has previously validated for its strong performance in echoic modal KWS. 

Our initial approach involved reducing the number of channels within the convolutional layers, referred to as width of the network. This adjustment was motivated by the fact that ResNet-18 was originally designed for tasks such as image classification and other computer vision applications, where visual data typically carries a richer information load compared to audio data. Consequently, reducing the network's width was a deliberate strategy aimed at decreasing the model's parameter count and mitigating the risk of overfitting. In our study, we explored width reduction in comparison to the original ResNet-18, as well as versions with widths reduced to 1/2, 1/4, and 1/8 of the original width. We assessed their respective performance within the echoic KWS pipeline.

Furthermore, we employed depthwise separable convolutions to replace convolution modules in the model. Depthwise separable convolution modules break down standard convolutions into depthwise convolution and pointwise convolution, significantly reducing the number of parameters and computations while maintaining similar performance. We compared the accuracy and parameter count differences when using depthwise separable convolution modules with different widths.

Our experiments were conducted on approximately 1800 speech samples from five participants, with a data split of 80\% for training and 20\% for testing. Random seeds were used to control data splitting and data augmentation parameters. For each model variant, we systematically adjusted the random seeds and conducted ten rounds of training and testing to obtain a reliable average accuracy. Furthermore, we employed \textit{torchstat} to analyze the model's parameters and computational complexity.

In the training process, we utilized the SGD optimizer with a learning rate following warm-up strategy, starting from 0 and linearly increasing to 0.1 over the first 50 epochs. Subsequently, the training continues for 1000 epochs with cosine learning rate decay. Additionally, we applied data augmentation including random noise, random padding, and overlaying background noise data.

\subsubsection{Results and Discussion}

The results are presented in \cref{tab:acc-param}, illustrating eight different model configurations resulting from combinations of four distinct network widths and the inclusion of depth-wise separable convolutions (DS). This table provides insights into the average accuracy, parameter count, and computational complexity of these models.

In the model naming convention, the fraction following \textit{ResNet-18} denotes the reduction in model width relative to the original model, and the suffix \textit{DS} indicates the utilization of depth-wise separable convolutions.

Notably, we observed that transitioning from ResNet-18 to ResNet-18-1/4-DS resulted in only a marginal 0.91\% increase in average WER, while substantially reducing the parameter count by over 100 times. This highlights the potential for our models to be efficiently deployed on XR headsets.

\begin{table}[ht]
\centering
\caption{Comparison of different models' performance, including Word Error Rate, Parameters, and Multiply-Add Operations.}
\label{tab:acc-param}
\begin{tabular}{lcrr}
\toprule
Model             & WER & Params & MAdd \\ \midrule
ResNet-18        & $5.06\% \pm 0.650$          & 11.22M     & 1.84G       \\
ResNet-18-1/2    & $5.47\% \pm 0.466$          & 2.820M     & 468.4M      \\
ResNet-18-1/4    & $5.81\% \pm 0.639$          & 712.6K     & 121.11M      \\
ResNet-18-1/8    & $7.69\% \pm 0.528$          & 182.0K     & 32.28M      \\
ResNet-18-DS     & $5.33\% \pm 0.539$          & 1.484M     & 264.1M      \\
ResNet-18-1/2-DS & $5.56\% \pm 0.714$          & 394.0K     & 79.99M      \\
ResNet-18-1/4-DS & $5.97\% \pm 0.573$          & 109.9K     & 24.74M      \\
ResNet-18-1/8-DS & $8.00\% \pm 1.260$          & 33.22K     & 8.93M      \\ \bottomrule
\end{tabular}
\end{table}

\subsection{Experiment 2: Noisy Environment}
\label{sec:experiment 2}

In this section, we examined the performance of two fusion strategies across different environments and signal-to-noise ratios (SNR). The research findings indicate that, in all experimental conditions, the performance of the Keyword Spotting (KWS) system using Vocal-Echoic dual-modal fusion surpasses that of its single-modal counterparts.

\subsubsection{Experiment Setup}

In this experiment, we simulated diverse scenarios by superimposing data with noise of varying intensities corresponding to specific scenes. We compared the average performance of different modalities within these scenarios to evaluate their effectiveness.

The signal-to-noise ratio (SNR) was employed to straightforwardly measure the strength of noise, defined as the ratio of signal power $P_s$ to noise power $P_n$. Decibels (dB) were utilized as the unit of measurement for SNR, as shown in the following formula:

\begin{equation}
SNR (dB) = 10\log_{10}{P_s/P_n}
\end{equation}

We investigated the performance of the fusion modality in environments with SNR ranging from $-10dB$ to $10dB$. The noise signals were multiplied by coefficients, which were determined based on the average power of the data and the target SNR, and then added to the data to simulate various intensity levels of noisy environments. The data were then subjected to vocal and echoic modality processing steps, as previously described, to generate predictive results. Subsequently, the fusion modality produced fused results. We compared the word error rates of these results to assess the performance of different modalities in these noisy environments.

\subsubsection{Results and Discussion}

\begin{figure}[tb]
 \centering
 \includegraphics[width=\linewidth]{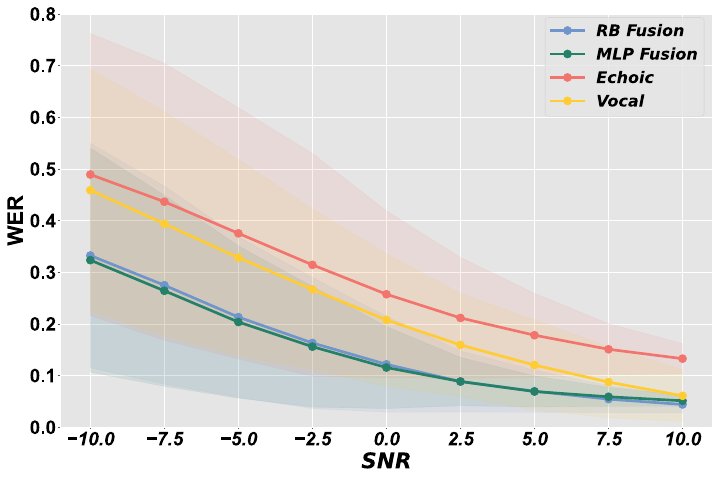}
 \caption{Comparison of average Word Error Rates (WER) between single-modal and dual-modal KWS systems in all noise scenarios. Our dual-modal systems (RB Fusion and MLP Fusion) achieve lower WER than single-modal systems (Echoic and Vocal) across all SNRs. At the strongest noise level (SNR=-10.0), MLP fusion reduces WER by 15.68\% and 16.57\% compared to vocal and echoic systems.}
 \label{fig:exp-2}
\end{figure}

In \cref{fig:exp-2}, we present the average results of two single modalities: vocal and echoic, and two dual-modal fusion strategies: reliability-based fusion (RB fusion), and MLP-based fusion (MLP fusion) across all scenarios. The standard deviations of these data are also depicted with shaded regions of the same color on the graph. Experimental results show that both MLP fusion and RB fusion outperform the two individual modalities.

\begin{figure*}[tb]
 \centering
 \includegraphics[width=\linewidth]{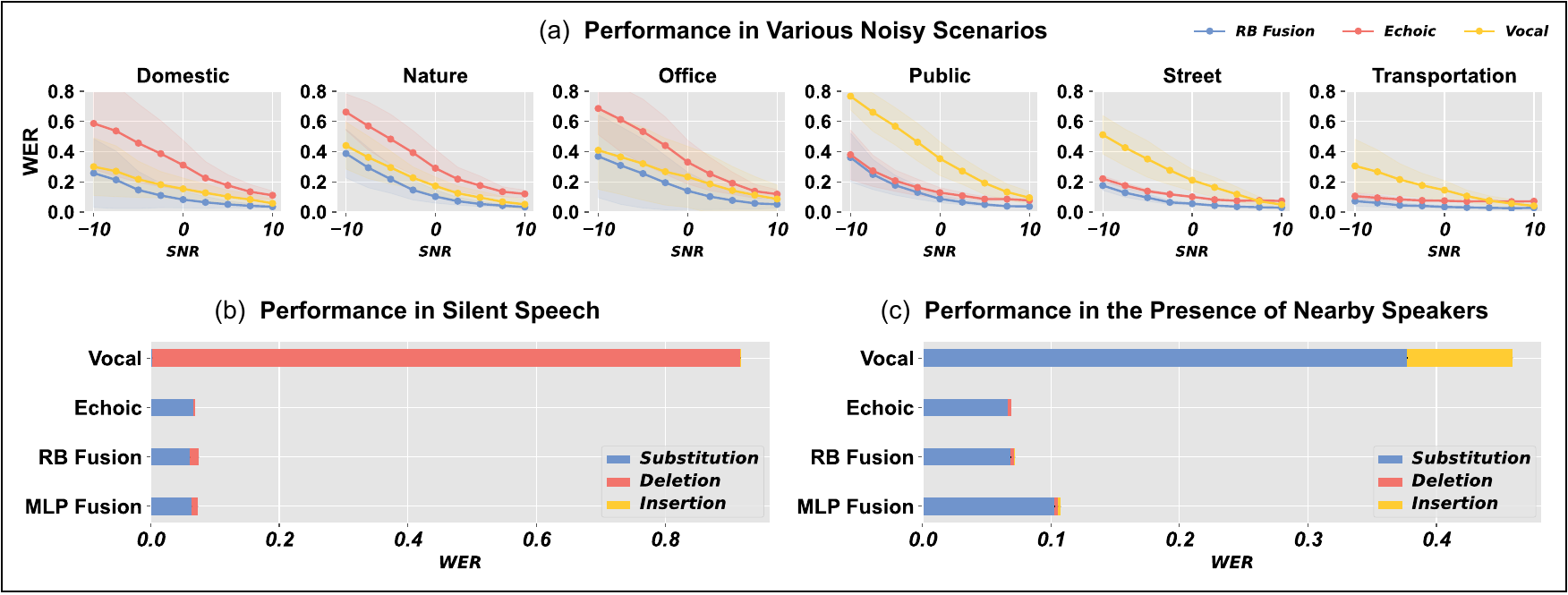}
 \caption{The performance of our system in three challenging scenarios, measured by Word Error Rate (WER). (a) Fusion consistently achieves the lowest WER, outperforming single modalities in all scenarios. (b) In silent speech, traditional vocal KWS fails entirely, while fusion matches the performance of the echoic modality, notably expanding the system's usage scenarios. (c) In the presence of nearby speakers, the fusion's WER is significantly lower than that of vocal modal systems, greatly reducing false triggers in speech-interference environments.}
 \label{fig:exp-1}
\end{figure*}

To provide a more detailed assessment of the system's performance across different modalities, we illustrate the experimental results of RB fusion in six distinct scenarios in \cref{fig:exp-1} (a). RB fusion exhibits slightly higher average word error rates, making it a more challenging test of fusion strategy performance. The results for each scenario are calculated as the average of sub-scenario experimental outcomes. In the first three scenarios, such as the Domestic setting with pronounced high-frequency noise interference, the vocal modality exhibits superior performance over the echoic modality. Conversely, in the latter three scenarios, such as the Public setting where noise primarily manifests as low-frequency disturbances, the echoic modality demonstrates a performance advantage over the vocal modality. Regardless of the individual modalities performance, RB fusion consistently exhibits lower word error rates than either of them in any given environment.

The experiments have shown that multimodal fusion consistently outperforms its single-modal counterparts in the majority of environmental conditions, confirming its remarkable robustness. It combines the advantages of both modalities and can be applied in a wider range of environments.

\begin{figure}[tb]
 \centering %
 \includegraphics[width=\linewidth]{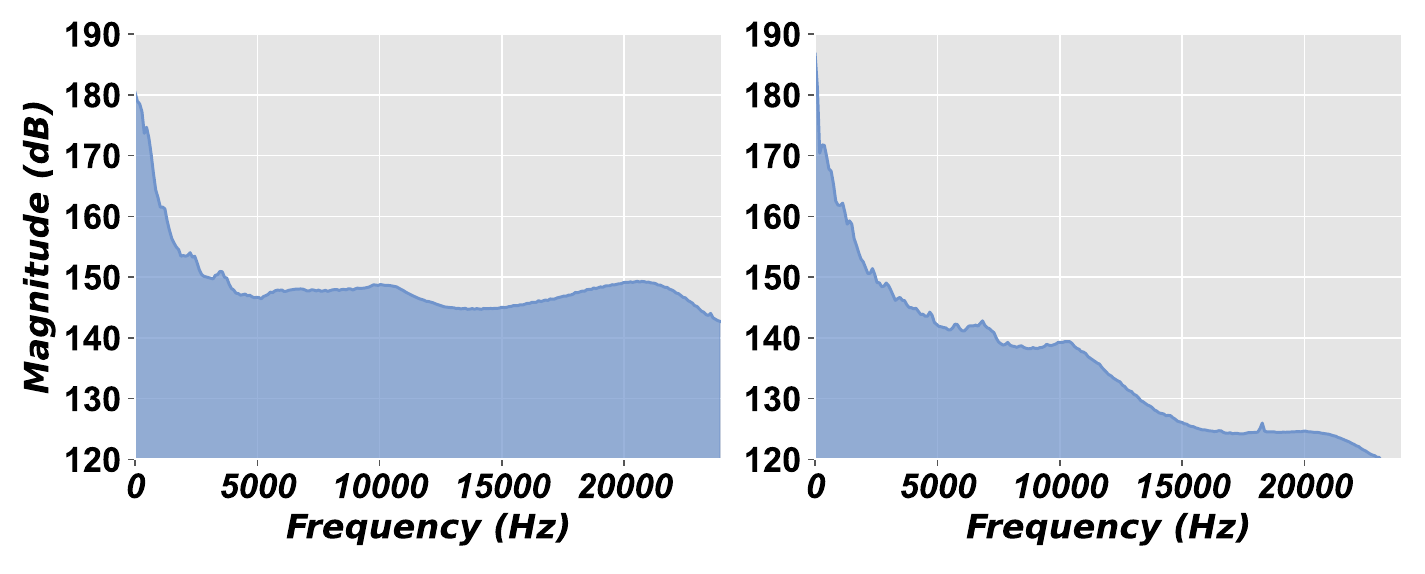}
 \caption{Frequency distribution of Meeting (left) and Metro (right). The frequency distribution of noise varies across different scenarios, with some noise having a significant presence in high frequencies, while another portion is primarily concentrated in lower frequencies.}
 \label{fig:noise-freq}
\end{figure}

\subsection{Experiment 3: Silent Speech}

In this portion of the experiment, we conducted a comparison between two fusion strategies and two single-modal approaches in an echoic-only state. Research findings suggest that all fusion strategy can effectively utilize the information provided by the echoic modality, resulting in a significantly lower word error rate compared to using the vocal modality alone. This supports the capacity of our designed dual-modal Keyword Spotting (KWS) system to provide dependable results in environments where user silence is necessary.

\subsubsection{Experiment Setup}

In this assessment, we employed data from which vocal information had been removed to evaluate the performance of the fusion modality in a silent environment.

Because the experiment relies on echoic information, we continued to use the self-recorded dataset mentioned earlier. We applied the previously mentioned filtering and separation process to all test data, categorizing it into vocal and echoic signals. The vocal segments were substituted with corresponding portions of random silence signals. In this scenario, vocal information is entirely eliminated.

Following that, the echoic and vocal signals are separately fed into their corresponding modalities, resulting in predictions. The fusion modality will generate results based on these predictions and will be compared to both the echoic and vocal modalities to evaluate the system's performance in a silence environment.

\subsubsection{Results and Discussion}

The experimental results, shown in \cref{fig:exp-1} (b), display the Word Error Rates (WER) for each modality in the form of bar charts. Various colors on the bars represent the proportions of substitutions, deletions, and insertions. As expected, with the exception of segments in the test set that originally contained silence signals, all signals have been transformed into deletions in the vocal modality, effectively eliminating the possibility of vocal signals providing information. At the same time, the performance of the echoic modality remains unaffected, with error recognition primarily consisting of substitutions and deletions, which aligns with our expectations.

The results indicate that the Word Error Rate (WER) of the two fusion strategies closely approximates that of the echoic modality, which is significantly lower than that of the vocal modality. 

In conclusion, the experimental results showcase that our fusion strategy can yield accurate results in an echoic-only environment, a capability that cannot be achieved by the vocal-only single-modal approach. This signifies that our device can operate without requiring additional user intervention, allowing users the freedom to choose whether to speak or use silent speech, and delivers reliable results.

\subsection{Experiment 4: Nearby Speaker Interference}

\label{sec:experiment 4}
In the following analysis, we conducted an investigation into the performance of fusion methods with vocal signals in the presence of interference from other vocal sources. Research findings indicate that the fusion strategy can effectively harness the information from echoic signals. Ultimately, the accuracy of the fusion strategy is within an acceptable range, slightly higher than the performance level of the pure echoic modality and far below that of the vocal modality. This discovery attests that our system can exhibit noteworthy performance even in the face of substantial interference from vocal noise, underscoring its robustness.

\subsubsection{Experiment Setup}

During this experiment, we simulated the presence of nearby speakers by superimposing the voices of other individuals onto the data, enabling an examination of the system's robustness under these conditions. Only the vocal component was contaminated; the echoic component remained unaffected. We employed the previously described self-recorded dataset for our experimentation, with the vocal data used to introduce interference sourced from the Google Speech Commands v2 dataset\cite{warden_speech_2018}. Accounting for differences in average power between the two datasets vocal components, we applied a fixed scaling factor to the vocal signals employed for superimposition to more accurately simulate nearby speakers.

Data from the self-recorded dataset, when superimposed with other human vocal signals, will yield results separately for the silent and vocal single modalities. Predictions from the fusion modality, based on these outcomes, will be compared to assess the system's performance in scenarios involving nearby speakers.

\subsubsection{Results and Discussion}

The results of this experiment are illustrated in \cref{fig:exp-1} (c), where error rates for each modality are presented in the form of bar charts. Different colors of bars represent the proportions of substitution, deletion, and insertion errors. In this context, the vocal modality displayed nearly half of the errors, primarily composed of substitution and deletion errors. In contrast, the echoic modality still produced relatively reliable results, with a certain proportion of substitution errors and very few deletion errors. From the results, it is evident that the fusion modality significantly alleviated the high error rate observed in the vocal modality, including both insertion and substitutions errors.

This experiment demonstrates that even in situations characterized by highly conspicuous background human speech interference, the fusion strategy can effectively prevent false triggers and provide reliable responses. The significant reduction in substitution errors suggests that users can have confidence in the recognition accuracy, even in the presence of human voice noise. Additionally, the near elimination of insertion errors addresses concerns related to unintentional activations in noisy or crowded settings. Users can confidently utilize this approach in environments such as discussions, classrooms, and similar settings without concern for the impact of high-decibel human voices in these scenarios.

\subsection{Discussion}

Our experimental results demonstrate that our dual-modal approach is more robust compared to both vocal keyword spotting methods and silent speech methods:

\textbf{Comparison with Vocal KWS method.} Our dual-modal approach demonstrates superior performance in various noisy environments compared to conventional keyword spotting methods. Additionally, it can effectively filter out interference from other nearby speakers and supports the use of silent speech when vocalization is inconvenient.
                
\textbf{Comparison with Silent Speech method.} Our dual-modal approach inherits the capability of prior works on silent speech interfaces. Moreover, in scenarios where silent speech recognition performance experiences significant degradation, such as excessively noisy environments, intense physical activities, and ultrasonic interference, our system can seamlessly leverage vocal speech for recognition. This capability not only aids in avoiding potential system failures but also eliminates the need for manual mode switching.

As a supplement, we have developed a game to validate the experimental results, showcasing the effectiveness of our system in various scenarios. Demonstration video can be viewed at \url{https://youtu.be/fSQoEJ37uEw}.

\section{Limitations and Future Work}

Our work also has some limitations, and we discuss the limitations we have determined through additional studies in this section.

\textbf{Physical Activities.} Walking and head shaking impact the performance of both the vocal and echoic modality pipelines, consequently affecting the overall system. To investigate the influence of these factors, we conducted a series of experiments. The results indicate that: 1) Walking and head shaking have an observable but minor impact on the vocal modality. However, they significantly impact the performance of the echoic modality, causing an increase of around 50\% and around 33\% in WER, respectively. 2) The fusion method optimized with activities data can mitigate the interference from the echoic modality, resulting in fused WERs that are still lower than the individual WERs of the vocal modality. Nevertheless, physical activities still weaken our system's advantage over conventional vocal KWS. Further enhancement of the echoic modality method remains a potential direction for improvement.

\textbf{Ultrasonic Interference.} Another factor not directly addressed in this paper is the interference from ultrasound emitted by other devices of the same kind. The experiments indicate that: 1) The interference from a stationary ultrasound source is neglectable across multiple directions and distances. This may be because the differential processing in our echoic modality pipeline has a mitigating effect on ultrasound interference. 2) On the other hand, moving ultrasound source has a more significant impact, with an average increase of approximately 7\% in WER at a distance of 1m, and closer ultrasound sources causing greater interference. This is because motion diminishes the mitigating effect of the differential processing on interference. Further research into this aspect is a potential avenue for future work.

Other potential future work includes investigating the impact on the performance and user experience of integrating additional modalities in KWS, improving the performance of the echoic modality model in the system, and further minimizing the introduction of additional hardware.

\section{Conclusion}

In this study, we introduce a dual-modal keyword spotting (KWS) system for XR headsets, implemented on the Microsoft HoloLens 2 platform. The key of our system lies in the fusion of features from two distinct modalities: vocal speech and mouth movement information captured through ultrasonic echoes. This integration imparts superior noise robustness and adaptability to diverse scenarios. Specifically, our approach efficiently utilizes hardware, requiring only off-the-shelf speakers and microphones to obtain information from both modalities. Moreover, our method is computationally lightweight, employing streamlined models and efficient fusion strategies. Our experimental results demonstrate the exceptional performance of this dual-modal system across various challenging scenarios. It outperforms single-modal systems in noisy environments and offers advantages in silent scenarios and situations with nearby speech interference, where traditional vocal KWS systems struggle. Overall, our proposed dual-modal method enhances the noise robustness of KWS systems and notably expands their application scope. This advancement empowers users to engage in speech interactions more frequently, providing not only superior interaction experiences but also enhanced flexibility of choice.

\bibliographystyle{abbrv-doi-hyperref}

\bibliography{template}

\end{document}